\documentclass[11pt]{article}
\usepackage{mathrsfs}
\usepackage{amsfonts}
\usepackage{amsfonts}
\usepackage{mathrsfs}
\usepackage{amssymb}
\usepackage{hyperref,color}
\usepackage{exscale,relsize,graphics}

\usepackage{amsmath,amsthm}
\numberwithin{equation}{section}

\allowdisplaybreaks[1]

\author{\bf Yu Xiao$^1$,~Engui Fan$^1
\footnote{Corresponding
author and  e-mail address:
      faneg@fudan.edu.cn}$}
\date{ \small{$^1$ School of Mathematical Science, Fudan University, Shanghai
200433, P.R. China}}
\title{\bf A Riemann-Hilbert approach to the Harry-Dym equation on the line}

\begin{document}
\maketitle

\baselineskip=18pt

\begin{abstract}

\baselineskip=18pt

In this paper, we consider the  Harry-Dym equation on the line with decaying initial value. The Fokas unified method is used to construct  the solution  of the  Harry-Dym equation
 via  a $2 \times 2$ matrix Riemann Hilbert problem in the complex  plane.  Further,  one-cups soltion solution  is expressed   in terms of  solutions of the Riemann Hilbert problem.
\end{abstract}
\section{Introduction}
The following nonlinear partial differential equation
\begin{equation}\label{1.1}
  q_t - 2(\frac{1}{\sqrt{1+q}})_{xxx} = 0
\end{equation}
is known as the Harry-Dym equation \cite{W-B-C}. This equation was obtained by Harry-Dym and
Martin Kruskal as an evolution equation solvable by a spectral problem based on the string
equation instead of the Schro¡§dinger equation.
The Harry-Dym equation has interest in the study of the Saffman-Taylor problem which describes the motion of a two-dimensional interface between a viscous and a nonviscous fluid \cite{Kadanoff}.
The Harry-Dym equation shares many of the properties typical of
the soliton equations.   It is a completely integrable equation which can be solved by the inverse
scattering transform\cite{W}.  It has a bi-Hamiltonian structure \cite{M},   an infinite number of conservation
laws and infinitely many symmetries \cite{L-L},  and  has  reciprocal Backlund transformations to  the KdV equation  \cite{R-N}.
The  Harry-Dym equation has been solved  in different method such as the inversing scatting method \cite{W},  the B$\ddot{a}$cklund transformation technique \cite{LLS},  the straightforward method \cite{B-D-H}.   Especially,   the Wadati obtained  the one-cups soliton solution  \cite{W}
$$q(x,t)=\tanh^{-4}(\kappa x-4\kappa^{3}t+\kappa x_{0}+\varepsilon_{+})-1,$$
 $$\varepsilon_{+}=\frac{1}{\kappa}[1+\tanh(\kappa x-4\kappa^{3}t+\kappa x_{0}+\varepsilon_{+})].$$
  by using inverse scattering transformation.

The main aim of this paper is to develop the inversing scatting method, based on an Riemann-Hibert problem for solving  nonlinear integrable systems called
 unified  method \cite{F1997} which has been further  developed and  applied in different   equations  with  initial value problems on the line \cite{F, F2008, L-F2,L-F3,L} and  initial boundary value problem  on  half line \cite{L-F, M-S, M-S1}.   In this paper,   we consider the initial value problem of the Harry-Dym equation
\begin{equation}\label{1.1}
  q_t - 2(\frac{1}{\sqrt{1+q}})_{xxx} = 0,\  x\in R, \ t>0,
\end{equation}
$$
q(x,0)=q_0(x),
$$
where the $q_0(x)$ is a smoothly real-valued function and  decay as $|x|\rightarrow\infty$.
 The organization of the paper is as follows.   In the following section 2,  we perform the spectral analysis of the associated Lax pair
 for the Harry-Dym equation.  In section 3,   we formulate the main Riemann-Hilbert problem associated with the initial value problem (\ref{1.1}).
 In section 4, we obtain one-cups soltion solution in the terms of  Riemann-Hilbert
problem, which has a  similar, but not the same form  constructed  by the inverse scatting method \cite{W}.

\section{Spectral analysis }\label{section2}
{\bf 2.1 A Lax pair}

In general, the matrix Riemann Hilbert problem is defined in the $\lambda$ plane and  has explicit $(x,t)$ dependence,
 while for the Harry-Dym equation (\ref{1.1}),  we need to  construct a new matrix Riemann Hilbert problem with  explicit $(y,t)$ dependence, where  $y(x,t)$ is a function which is an unknown from the initial value condition.  For this purpose, we  make a   transformation
  $$ \rho=\sqrt{1+q},$$
 the equation (\ref{1.1}) can be expressed by
$$(\rho^{2})_{t}-2(\frac{1}{\rho})_{xxx}=0.$$
 Then the initial value problem   (\ref{1.1}) is
transformed into
 \begin{equation}\label{1.2}
  (\rho^{2})_{t}-2(\frac{1}{\rho})_{xxx}=0, x\in R,t>0,
\end{equation}
$$
\rho(x,0)=\rho_0(x)=\sqrt{1+q_0(x)},
$$
$$
\rho_0(x)\rightarrow 1,|x|\rightarrow\infty.
$$

It was shown that the equation (\ref{1.1}) admits
the following Lax pair \cite{W}
\begin{equation}\label{2.1}
\begin{cases}
    & \psi_{xx} = -\lambda^{2}(1+q)\psi, \\
    & \psi_t=2\lambda^{2}[\frac{2}{\sqrt{1+q}}\psi_{x}-(\frac{1}{\sqrt{1+q}})_{x}\psi].
\end{cases}
\end{equation}
Making a transformation
$$ \rho=\sqrt{1+q}, \ \varphi=\begin{pmatrix} \psi \\
 \psi_{x} \end{pmatrix},$$  then the Lax pair (\ref{2.1}) can be written in matrix form
\begin{equation}\label{2.2}
\begin{cases}
    & \varphi_x = M\varphi, \\
    & \varphi_t=N\varphi,
\end{cases}
\end{equation}
where
\begin{align*}
M = \begin{pmatrix} 0	&	1 \\
-\lambda^{2}\rho^{2}	&	0\end{pmatrix},\qquad
N =\begin{pmatrix} -2\lambda^{2}(\frac{1}{\rho})_{x} 		&	4\lambda^{2}\frac{1}{\rho}	\\
-4\lambda^{4}\rho-2\lambda^{2}(\frac{1}{\rho})_{xx}	&	2\lambda^{2}(\frac{1}{\rho})_{x}	\end{pmatrix}.
\end{align*}
Further by the gauge transformations
$$
\phi=\begin{pmatrix} 1	&	i \\
i	&	1\end{pmatrix}	
\begin{pmatrix} \sqrt{\lambda\rho}		&	0	\\
0	&	\frac{1}{\sqrt{\lambda\rho}}	\end{pmatrix}
\varphi.
$$
we have
\begin{equation}\label{2.3}
\begin{cases}
    & \phi_x +i\lambda\rho\sigma_{3}\phi= U\phi, \\
    & \phi_t+i(\lambda\frac{1}{\rho}(\frac{1}{\rho})_{xx}+4\lambda^{3})\sigma_{3}\phi=V\phi,
\end{cases}
\end{equation}
where
\begin{align*}
U(x,t) =\frac{1}{2}\frac{\rho_{x}}{\rho} \sigma_{2},
	\ \	
V(x,t,\lambda) =-\lambda\frac{1}{\rho}(\frac{1}{\rho})_{xx} \sigma_{1}-2\lambda^{2}(\frac{1}{\rho})_{x} \sigma_{2}.
\end{align*}
$$
\sigma_{1} =\begin{pmatrix} 0  & 1    \\
1  &   0 \end{pmatrix},\qquad \sigma_2 = \begin{pmatrix} 0 & -i \\ i & 0 \end{pmatrix},
 \qquad \sigma_3 = \begin{pmatrix} 1 & 0 \\ 0 & -1 \end{pmatrix}.
$$

It is clear that as $|x|\rightarrow\infty$, $U(x,t)\rightarrow 0,V(x,t,\lambda)\rightarrow 0$.
We define a real-valued function $y(x,t)$ by
$$
y(x,t)=x+\int^{\infty}_{x}(1-\rho(\xi,t))d\xi.
$$
It is obvious that
$$
y_{x}=\rho(x,t),\qquad y_{t}=-\int^{\infty}_{x}\rho_{t}(\xi,t))d\xi.
$$
The conservation law
$$
\rho_{t}-(-\frac{1}{2}((\frac{1}{\rho})_{x})^{2} +\frac{1}{\rho}(\frac{1}{\rho})_{xx})_{x}=0
$$
implies that
$$y_{t}=-\frac{1}{2}((\frac{1}{\rho})_{x})^{2} +\frac{1}{\rho}(\frac{1}{\rho})_{xx}.$$
Extending the column vector $\phi$ to be a $2\times2$ matrix and letting
$$\mu=\phi \mathrm{exp}(i\lambda y(x,t)\sigma_{3}+4i\lambda^{3}t\sigma_{3}),$$
then $\mu$ solves

\begin{equation}\label{2.4}
\begin{cases}
    & \mu_x +i\lambda y_{x}[\sigma_{3},\mu]= \widetilde{U}\mu, \\
    & \mu_t+i(\lambda y_{t}+4\lambda^{3})[\sigma_{3},\mu]=\widetilde{V}\mu,
\end{cases}
\end{equation}
which can be written in full derivative form
\begin{equation*}
d(e^{i(y(x,t) x+4\lambda^{3} t)\hat{\sigma_{3}}}\mu)=e^{i(y(x,t) x+4\lambda^{3} t)\hat {\sigma_{3}}}(\widetilde{U}dx+\widetilde{V}dt)\mu,
\end{equation*}
where
$$\widetilde{U}=U,$$
$$
\widetilde{V}=-\frac{1}{2}i\lambda((\frac{1}{\rho})_{x})^{2}\sigma_{3}
-\lambda \frac{1}{\rho}(\frac{1}{\rho})_{xx}\sigma_{1}-2\lambda^{2}(\frac{1}{\rho})_{x}\sigma_{2},
$$
$[\sigma_{3},\mu]=\sigma_{3}\mu-\mu\sigma_{3}$. As $|x|\rightarrow\infty$, $\widetilde{V}\rightarrow 0$. The
lax pair in the (\ref{2.4}) is very convenient for dedicated solutions via
integral Volterra equation, which is also what we study in the following paper.

{\bf  Remark 2.1} By the representation of $M,N$ and $U,V$ in (\ref{2.2}) and (\ref{2.3}) respectively, we find that
$\psi_{x},\psi_{t}$ and $\phi_{x},\phi_{t}$ have no singularity in $\lambda=0$. Therefore, $\phi$ has no real  singularity in $\lambda=0$.

{\bf 2.2 Eigenfunctions}
 We define two eigenfunctions $\mu_{\pm}$ of equation (\ref{2.4}) as the
 solutions of the following two Volterra integral equation in the $(x,t)$ plane
 \begin{equation}\label{2.5}
\mu(x,t,\lambda)=I+\int_{(x^{*}{},t^{*})}^{(x,t)}e^{-[i\lambda( y(x,t)-y(x',t))+4i\lambda^{3} (t-\tau)]\hat{\sigma_{3}}}
(\widetilde{U}(x',t)\mu(x',t,\lambda)dx'+\widetilde{V}(x',\tau,\lambda)\mu(x',\tau,\lambda))d\tau
\end{equation}
 where $I$ is the $ 2\times2$ identity matrix, $\hat{\sigma_{3}}$ acts on  a $2\times2$ matrix $A$
 by $\hat{\sigma_{3}}A=\sigma_{3}A\sigma_{3}$. Since
 the integrated expression  is independent of the path of integration, we choose the particular initial points of integration to be
 parallel to the $x$ -axis and obtain that
 $\mu_{+}$ and $\mu_{-}$

$$
\mu_{+}(x,t,\lambda)=I-\int_{x}^{\infty}e^{-i\lambda( y(x,t)-y(x',t))\hat{\sigma_{3}}}\widetilde{U}(x',t)\mu_{+}(x',t,\lambda)dx',
$$
\begin{equation}\label{2.6}
\mu_{-}(x,t,\lambda)=I+\int_{-\infty}^{x}e^{-i\lambda( y(x,t)-y(x',t))\hat{\sigma_{3}}}\widetilde{U}(x',t)\mu_{-}(x',t,\lambda)dx'.
\end{equation}
Define the following sets
$$
D_1 = {\{{\lambda \in C| {\mathrm{Im}\lambda>0}}}\},
$$
$$
D_2 = {\{{\lambda \in C| {\mathrm{Im}\lambda<0}}}\}.
$$
Since any fixed $t$, $y_{x}=\rho(x,t)>0$, $y(x,t)$ is increasing function of $x$ for fixed $t$.
 as $x-x'<0$, $y(x,t)-y(x',t)<0$; as $x-x'>0$, $y(x,t)-y(x',t)>0$. We can deduce that the
 second column vectors of $\mu_{+},\mu_{-}$ are bounded and analytic for $\lambda \in C$ provided
 that $\lambda$ belongs to $D_1,D_2$, respectively. We denote these vectors with superscripts
 (1),(2) to indicate the domains of their boundedness. Then
 $$
 \mu_+=(\mu_+^{(2)},\mu_+^{(1)}),\mu_-=(\mu_-^{(1)},\mu_-^{(2)}).
 $$
For any $x,t$, the following conditions are satisfied
$$
(\mu_-^{(1)},\mu_+^{(1)}) =I+O(1/\lambda), \lambda\rightarrow\infty, \lambda\in D_1,
$$
$$
(\mu_+^{(2)},\mu_-^{(2)}) =I+O(1/\lambda), \lambda\rightarrow\infty, \lambda\in D_2,
$$
$$
\mu_\pm=I+O(1/\lambda), \lambda\rightarrow\infty.
$$

{\bf 2.3 Spectral functions} For $\lambda\in R$, the eigenfunction $\mu_+,\mu_-$ being the solution of the system
of differential equation (\ref{2.4}) are related by a matrix independent of $(x,t)$. We
define the spectral function by
\begin{equation}\label{2.7}
\mu_{+}(x,t,\lambda)=\mu_{-}(x,t,\lambda)e^{-i(\lambda y(x,t)+4\lambda^{3} t)\hat{\sigma_{3}}}s(\lambda).
\end{equation}
From (\ref{2.4}), we get
\begin{equation}\label{2.8}
det(\mu_\pm(x,t,\lambda))=1.
\end{equation}
Since $\overline{\widetilde{U}}(x,t)=-\widetilde{U}(x,t)$, the $\mu_\pm(x,t,\lambda)$ have the relations
\begin{equation}\label{2.9}
\begin{cases}
    &\mu_{\pm11}(x,t,\lambda) = \overline{\mu_{\pm22}(x,t,\bar{\lambda})},\quad \mu_{\pm21}(x,t,\lambda) = \overline{\mu_{\pm12}(x,t, \bar{\lambda})},\\
    & \mu_{\pm11}(x, t, -\lambda) = \mu_{\pm22}(x, t, \lambda),\quad \mu_{\pm12}(x, t, -\lambda) = \mu_{\pm21}(x, t,\lambda) .
\end{cases}
\end{equation}
The spectral function $s(\lambda)$ can be written as
\begin{equation}\label{2.10}
s(\lambda) = \begin{pmatrix} \overline{a(\bar{\lambda})} & b(\lambda) \\
 \overline{b(\bar{\lambda})}  &   a(\lambda) \end{pmatrix},
\end{equation}

\begin{equation}\label{2.11}
s(\lambda) =I-\int_{-\infty}^{+\infty}e^{i\lambda y(x,0)\hat{\sigma_{3}}}\widetilde{U}(x',0)\mu_{+}(x',0,\lambda)dx',\quad Im\lambda=0.
\end{equation}
From the (\ref{2.8}), $det(s(\lambda))=1$.
Equation (\ref{2.7}) and (\ref{2.8}) imply $a(\lambda)$ and $b(\lambda)$ have the
following properties:
\begin{itemize}
\item   $a(\lambda)$ is analytic in $D_1$ and  continuous  for $\lambda \in \bar{D}_1$ .
\item  $b(\lambda)$ is continuous for $\lambda\in R$.
\item $a(\lambda)\overline{a(\bar{\lambda})} -  b(\lambda)\overline{b(\bar{\lambda})} = 1, \quad \lambda \in R$.
\item $a(\lambda) = 1 + O\left(\frac{1}{\lambda}\right), \quad \lambda \to \infty,\quad \lambda \in D_1$.
\item  $b(\lambda) = O\left(\frac{1}{\lambda}\right), \quad \lambda \to \infty,\quad \lambda \in R.$
\end{itemize}

{\bf 2.4 Residue conditions}
We assume that $a(\lambda)$ has $N$ simple zeros $\{\lambda_j\}_{j=1}^{N}$ in the upper half plane.
These eigenvalues are  purely imaginary.
The second column of equation (\ref{2.7}) is
\begin{equation}\label{2.12}
\mu_+^{(1)}=b(\lambda)\mu_-^{(1)}e^{-2i(\lambda y(x,t)+4\lambda^{3}t)}+\mu_-^{(2)}a(\lambda).
\end{equation}
For the (\ref{2.8})and equation (\ref{2.12}), it yields
$$
a(\lambda)=det (\mu_-^{(1)},\mu_+^{(1)})
$$
where we have used that both sides are well defined and analytic in $D_1$ to
extend the above relation to $\bar{D}_1$. Hence if $a(\lambda_{j})=0$, the
$\mu_-^{(1)},\mu_+^{(1)}$ are linearly dependent vectors for each $x$ and $t$, i.e.
there exist  constants $b_{j}\neq 0$ such that
$$
\mu_-^{(1)}=b_{j}e^{2i(\lambda_{j} y(x,t)+4\lambda_{j}^{3}t)}\mu_+^{(1)},x\in R,t>0.
$$
Recalling the symmetries in the (\ref{2.9}), we find
$$
\mu_-^{(2)}=\bar{b}_{j}e^{-2i(\bar{\lambda}_{j} y(x,t)+4\bar{\lambda}_{j}^{3}t)}\mu_+^{(2)},x\in R,t>0.
$$
Consequently, the residues of $\mu_-^{(1)}/a$ and $\mu_-^{(2)}/\overline{a(\bar{\lambda})}$
at $\lambda_{j}$ and $\bar{\lambda}_{j}$ are

\begin{align*}
\underset{\lambda=\lambda_j}{\text{Res}} \frac{\mu_-^{(1)}(x,t,\lambda)}{a(\lambda)}=& C_j e^{2i(\lambda_{j} y(x,t)+4\lambda_{j}^{3}t)}\mu_+^{(2)}(x,t,\lambda_{j}), \qquad j = 1, \dots, N,
        \\
\underset{k=\bar{\lambda}_j}{\text{Res}} \frac{\mu_-^{(2)}(x,t,\lambda)}{\overline{a(\bar{\lambda})}} = &
 \bar{C}_j e^{-2i(\bar{\lambda_{j}} y(x,t)
 +4\bar{\lambda}_{j}^{3}t)} \mu_+^{(1)}(x,t,\bar{\lambda}_{j}), \qquad j = 1, \dots, N,
\end{align*}
where  $C_j=\frac{b_{j}}{\dot{a}(k_{j})}$, $\dot{a}(k) = \frac{da}{dk}$.

{\bf Remark 2.2} There is  the relation of $\mu_{\pm}$ that  the  $s(\lambda)$ is the scattering
matrix for the one dimensional Sch$\ddot{o}$dinger equation

\begin{align*}
W_{yy}+\lambda^{2}W=f(y)W
\end{align*}
via the Liouville transformation:
$$
y=x+\int^{\infty}_{x}(1-\rho(\xi,0))d\xi,\qquad W(y,\lambda)=\psi(y,\lambda)\rho_{0}(y)
$$

$$
 \rho_{0}(y)=\rho_{0}(x), \qquad f(y)=\frac{1}{2}(\rho_{0yy}\rho_{0}^{-1}-\frac{1}{2}\rho_{0y}^{2}\rho_{0}^{-2}).
$$
Therefore, in terms of spectral problem of Schr\"odinger equation, we deduce that $a(\lambda)$ only has
pure imaginary of simple poles in the upper plane.

\section{The Riemann-Hilbert Problem }\label{section3}
{\bf 3.1 A Riemann-Hilbert problem for $(x,t)$}
We now apply uniform method to solve the initial value problem for equation (\ref{1.2}) on the line,
and the solution can be expressed in terms of a $2\times2$ matrix Riemann-Hilbert problem.
Let $M(x,t,\lambda)$ be defined by
\begin{align}\label{3.1}
M_+ = \left(\frac{\mu_-^{(1)}}{a(\lambda)}, \mu_+^{(1)}\right), \quad \lambda \in D_1; \quad
M_- = \left(\mu_+^{(2)}, \frac{\mu_-^{(2)}}{\overline{a(\bar{\lambda})}}\right), \quad \lambda \in D_2
\end{align}
and
the $M$ satisfied the jump condition:
\begin{equation*}
  M_+(x,t,\lambda) =  M_-(x,t, \lambda)J(x,t,\lambda), \quad Im\lambda=0,
\end{equation*}
where
\begin{equation}\label{3.2}
J(x,t,\lambda) = \begin{pmatrix} \frac{1}{a(\lambda)\overline{a(\bar{\lambda})}} &\frac{b(\lambda)}{\overline{a(\bar{\lambda})}}e^{-2i(\lambda y(x,t)+4\lambda^{3}t)} \\
  -\frac{ \overline{b(\bar{\lambda})}}{a(\lambda)}e^{2i(\lambda y(x,t)+4\lambda^{3}t)}  &   1\end{pmatrix}, \quad Im\lambda=0.
\end{equation}
These definitions imply
\begin{equation}\label{3.3}
det M(x,t,\lambda)=1
\end{equation}
and
\begin{equation}\label{3.4}
 M(x,t,\lambda)=I+O(\frac{1}{\lambda}),\quad \lambda \rightarrow \infty.
\end{equation}
This contour of RH problem is the real axis.

 the jump matrix $J(x,t,\lambda)$ , the spectral $a(\lambda)$ and $b(\lambda)$
are dependent on the $ y(x,t)$, while $ y(x,t)$ is not involve
initial data. Therefore, this RH problem can't be formulated
in terms of initial alone. In order to overcome this problem,
we will reconstruct a new jump matrix by changing
 $$
 (x,t)\rightarrow (y,t),\quad y=y(x,t),
 $$
$y$ is a new scale. Then we can transform this RH problem into
the RH problem parametrized by $(y,t)$.

{\bf 3.2 A Riemann-Hilbert problem for $(y,t)$}

{\bf 3.1  The theorem}
Let $q_0(x),x\in R$ be a smooth function and decay as $|x|\rightarrow \infty$. Moreever
$1+q_0(x)>0$. Define the $\widetilde{U}_{0},\rho_{0}$ and $y_{0}(x)$ as follows:
$$
\widetilde{U}_{0}(x)=\frac{1}{2} \frac{\rho_{0x}(x)}{\rho_{0}(x)}\sigma_{2}, \quad\rho_{0}(x)=\sqrt{1+q_0(x)},
$$
$$
y_{0}(x)=x+\int^{\infty}_{x}(1-\rho_{0}(\xi))d\xi.
$$
Let $\mu_{+}(x,0,\lambda)$ and $\mu_{-}(x,0,\lambda)$ be the unique solution of
the Volterra linear integral equation (\ref{2.4}) evaluated at $t=0$ with
$\widetilde{U}_{0}(x,0)=\widetilde{U}_{0}(x)$,$\rho_{0}(x)=\rho(x,0)$ and $y_{0}(x)=y(x,0)$.
Define ${a(\lambda),b(\lambda),C_{j}}$ by
\begin{equation}\label{3.5}
\begin{pmatrix} b(\lambda) \\
 a(\lambda) \end{pmatrix} = [s(\lambda)]_2, \quad s(\lambda) =I-\int_{-\infty}^{+\infty}e^{i\lambda y_{0}(x)\hat{\sigma_{3}}}\widetilde{U}_{0}(x')\mu_{+}(x',0,\lambda)dx',\quad Im\lambda=0
\end{equation}
and
\begin{equation}\label{3.6}
[\mu_{-}(x,0,\lambda_{j})]_{1}=\dot{a}(\lambda_{j})C_j e^{2i\lambda_{j} y_{0}(x)}[\mu_+(x,0,\lambda_{j})]_{2}, \quad j = 1, \dots, N,
\end{equation}
where ($[A]_1$ $[A]_2$) denotes the first (second) column of a $2 \times 2$ matrix $A$.
We assume that $a(\lambda)$ has $N$ simple zeros $\{\lambda_j\}_{j=1}^{N}$ in the upper half plane and are pure imaginary. Then
\begin{itemize}
\item   $a(\lambda)$ is defined  for $k \in \bar{D}_1$  and  analytic in $D_1$ .
\item  $b(\lambda)$ is defined  for $\lambda\in R.$
\item $a(\lambda)\overline{a(\bar{\lambda})} -  b(\lambda)\overline{b(\bar{\lambda})} = 1, \quad \lambda \in R.$
\item $a(\lambda) = 1 + O\left(\frac{1}{\lambda}\right), \lambda \to \infty,\lambda \in D_1.$
\item  $b(\lambda) = O\left(\frac{1}{\lambda}\right),  \lambda \to \infty,\lambda \in R.$
\end{itemize}

Suppose there exists a uniquely solution $q(x,t)$ of equation (\ref{1.1}) with initial
data $q_{0}(x)$ such that $\rho_{0}(x)=\sqrt{1+q_{0}(x)}$ has sufficient smoothness and decay
for $t>0$. Then $q(x,t)$ is given in parametric form by
\begin{equation}\label{3.7}
q(x(y,t),t)=e^{8\int_{y}^{+\infty}m(y',t)dy'}-1
\end{equation}
and the function $x(y,t)$ is defined by
\begin{equation}\label{3.8}
x(y,t)=y+\int^{y}_{-\infty}(e^{-4\int^{\infty}_{y'}m(\xi,t)d\xi}-1)dy',
\end{equation}
where
$m(y,t) =-i\lim_{\lambda \to \infty} (\lambda M(y,t, \lambda))_{12},$
and $M(y,t, \lambda)$ is the uniquely solution of
the following RH problem
\begin{itemize}
\item $M(y,t,\lambda) = \left\{ \begin{array}{ll}
M_-(y,t,\lambda), &  \lambda \in D_2 , \\
M_+(y,t,\lambda), &  \lambda \in  D_1.\\
\end{array} \right.$

is a sectionally meromorphic function.

\item $M_+(y,t,\lambda) = M_-(y,t,\lambda) J^{(y)}(y,t,\lambda), \quad Im\lambda=0,$

where $J^{(y)}(y,t,\lambda)$ is defined by
\begin{equation}\label{3.9}
J^{(y)}(y,t,\lambda) = \begin{pmatrix} \frac{1}{a(\lambda)\overline{a(\bar{\lambda})}} &\frac{b(\lambda)}{\overline{a(\bar{\lambda})}}e^{-2i(\lambda y+4\lambda^{3}t)} \\
 - \frac{ \overline{b(\bar{\lambda})}}{a(\lambda)}e^{2i(\lambda y+4\lambda^{3}t)}  &   1\end{pmatrix}, \quad Im\lambda=0.
\end{equation}
\item
\begin{equation}\label{3.10}
M(y,t,\lambda) =I+ O\left(\frac{1}{\lambda}\right), \quad \lambda \to \infty.
\end{equation}
\item The possible simple poles of the first column of $M_+(y,t,\lambda)$ occur at $\lambda = \lambda_j$, $j = 1, \dots, N$, and the possible simple poles of the second column of $M_-(y,t,\lambda)$ occur at $\lambda = \bar{\lambda}_j$, $j = 1, \dots, N$.
The associated residues are given by
\begin{align}\label{3.11}
\underset{\lambda=\lambda_j}{\text{\upshape Res}} [M(y,t,\lambda)]_1 = C_{j}e^{2i(\lambda_{j}y+4 \lambda_{j}^{3}t)} [M(y,t,\lambda_{j})]_2,  \quad j = 1, \dots, N,
\\ \label{3.12}
\underset{\lambda=\bar{\lambda}_j}{\text{\upshape Res}} [M(y,t,\lambda)_2 = \bar{C}_{j} e^{-2i(\bar{\lambda}_{j} y+4\bar{\lambda}_{j}^{3}t)} [M(y,t,\bar{\lambda}_{j})]_1,  \quad j = 1, \dots, N.
\end{align}
\end{itemize}
Proof:
Assume that $\mu(x,t)$ is the solution of equation (\ref{2.4}), the asymptotic expansion of it
$$
\mu(x,t,\lambda)=I+\frac{\mu^{(1)}(x,t)}{\lambda}+\frac{\mu^{(2)}(x,t)}{\lambda^{2}}+\frac{\mu^{(3)}(x,t)}{\lambda^{3}}+O(\frac{1}{\lambda^{4}}),\lambda\rightarrow\infty
$$
into the $x$-part of equation (\ref{2.4}), where $\mu^{(1)}(x,t),\mu^{(2)}(x,t)$ and $\mu^{(3)}(x,t)$ are $2\times2$ matrixes, dependent on $x,t$.
 by considering the terms of $O(1)$, We get
\begin{equation}\label{3.13}
4\mu_{12}^{(1)}(x,t)=-\frac{\rho_{x}(x,t)}{\rho(x,t)}.
\end{equation}
By construction of the new RH problem about $(y,t,\lambda)$, we can deduce that
\begin{equation}\label{3.14}
\mu_{12}^{(1)}(x,t)=-i\lim_{\lambda \to \infty} (\lambda M(y,t, \lambda))_{12}=m(y,t).
\end{equation}
Then
\begin{equation}\label{3.15}
-\frac{1}{4}\frac{\rho_{x}(x,t)}{\rho(x,t)}=m(y,t).
\end{equation}
Equation (\ref{3.13}) can be expressed in terms of $y=y(x,t)$. Indeed, using
$\frac{dy}{dx}=\rho$, then (\ref{3.15}) becomes
\begin{equation}\label{3.16}
-\frac{1}{4}\frac{\rho_{y}}{\rho}=m(y,t).
\end{equation}
As $|y|\rightarrow \infty$, $\rho(y,t)\rightarrow 1$, by the evauation of (\ref{3.16}),
we get
$$
\rho(y,t)=e^{4\int_{y}^{+\infty}m(y',t)dy'}
$$
Therefore,
$$
q(x,t)=e^{8\int_{y}^{+\infty}m(y',t)dy'}-1
$$
As $|x|\rightarrow \infty$, $|y|\rightarrow \infty$ and $\frac{dy}{dx}=\rho>0$, so
$$
x=y+\int_{-\infty}^{y}(e^{-4\int_{y'}^{+\infty}m(\xi,t)d\xi}-1)dy'.
$$

{\bf  Remark 3.1}
It follows from the symmetries (\ref{2.9}) that the solution $M(y,t, \lambda)$
of Riemann Hilbert problem in the 3.1 theorem has the symmetries

\begin{equation}\label{3.17}
\begin{cases}
    & M_{11}(y,t, \lambda) = \overline{M_{22}(y,t,\bar{\lambda})},\quad  M_{21}(y,t, \lambda) = \overline{M_{12}(y,t, \bar{\lambda})}, \\
    &   M_{11}(y, t, -\lambda) = M_{22}(y, t, \lambda), \quad M_{12}(y, t, -\lambda) = M_{21}(y, t, \lambda).
\end{cases}
\end{equation}

\section{Soliton solution  }\label{section3}
The solitons correspond to spectral data ${\{a(\lambda),b(\lambda),C_{j}}\}$ for which
$b(\lambda)$ vanishes identically. In this case the jump matrix $J^{(y)}(y,t,\lambda)$
in the (\ref{3.9}) is the identity matrix and the RH problem of 3.1 theorem consists of
finding a meromorphic function $M(y,t, \lambda)$ satisfying (\ref{3.10}) and the residue
conditions (\ref{3.11}) and (\ref{3.12}). From (\ref{3.10})and (\ref{3.11}),
we get

\begin{align}\label{4.1}
 [M(y,t,\lambda)]_1 = \begin{pmatrix} 1 \\ 0 \end{pmatrix}+\sum\limits_{j=1}^{N}
\frac{C_{j}}{\lambda-\lambda_{j}}e^{2i(\lambda_{j}y+4 \lambda_{j}^{3}t)} [M(y,t,\lambda_{j})]_2.
\end{align}
for the symmetries (\ref{3.17}), equation (\ref{4.1}) can be written as
\begin{align}\label{4.2}
  \begin{pmatrix} \overline{M_{22}(y,t,\bar{\lambda}) }\\ \overline{M_{12}(y,t,\bar{\lambda})} \end{pmatrix} = \begin{pmatrix} 1 \\ 0 \end{pmatrix}+\sum\limits_{j=1}^{N}
\frac{C_{j}}{\lambda-\lambda_{j}}e^{2i(\lambda_{j}y+4 \lambda_{j}^{3}t)} \begin{pmatrix} M_{12}(y,t,\lambda_{j})\\ M_{22}(y,t,\lambda_{j}) \end{pmatrix}.
\end{align}
Evaluation at $\bar{\lambda}_{n}$ , equation (\ref{4.2}) becomes
\begin{align}\label{4.3}
  \begin{pmatrix} \overline{M_{22}(y,t,\lambda_{n}) }\\ \overline{M_{12}(y,t,\lambda_{n})} \end{pmatrix} = \begin{pmatrix} 1 \\ 0 \end{pmatrix}+\sum\limits_{j=1}^{N}
\frac{C_{j}}{\bar{\lambda}_{n}-\lambda_{j}}e^{2i(\lambda_{j}y+4 \lambda_{j}^{3}t)} \begin{pmatrix} M_{12}(y,t,\lambda_{j})\\ M_{22}(y,t,\lambda_{j}) \end{pmatrix},\quad n = 1, \dots, N.
\end{align}
Solving this algebraic system for $M_{12}(y,t,\lambda_{j}),M_{22}(y,t,\lambda_{j}),n = 1, \dots, N$, and
substituting them into (\ref{4.1}) provides a explicit expression for the $[M(y,t,\lambda)]_1$.
In terms of the symmetries (\ref{3.17}), we can get that $M_{12}(y,t,\lambda)$, which solves the Riemann
Hilbert problem. Then
 $$
 -i\lim_{\lambda \to \infty} (\lambda M(y,t, \lambda))_{12}=m(y,t)=-i\sum\limits_{j=1}^{N}C_{j}e^{2i(\lambda_{j}y+4 \lambda_{j}^{3}t)}  M_{12}(y,t,\lambda_{j}).
$$
Therefore, the $N$ soliton solution $q(x,t)$ is expressed by the (\ref{3.7}).

{\bf 4.1  One-soliton solution}

In this section we derive a explicit formulate for the one-soliton solution, which arise when $a(\lambda)$ has
a pure imaginary $\lambda_{1}$ of simple zero. Letting $N=1$ in (\ref{4.3}), from the the symmetries of
(\ref{2.9}), we can deduce that $a(\lambda_{1})=\overline{a(-\bar{\lambda}_{1})}=0$, then $\lambda_{1}=-\bar{\lambda}_{1}$ and $\dot{a}(\lambda_{1})=\overline{\dot{a}(-\bar{\lambda}_{1})}$. Since  the $b_{1}$ is a real constant, we find that $C_{1}=-\overline{C_{1}}$, thus $C_{1}$
is a pure  imaginary. Making use of the symmetries of
(\ref{3.17}), we can obtain
$$
\overline{M_{22}(y,t,\lambda_{1})}=1+
\frac{C_{1}}{\bar{\lambda}_{1}-\lambda_{1}}e^{2i(\lambda_{1}y+4 \lambda_{1}^{3}t)}M_{12}(y,t,\lambda_{1}),
$$
$$
\overline{M_{12}(y,t,\lambda_{1})}= \frac{C_{1}}{\bar{\lambda}_{1}-\lambda_{1}}e^{2i(\lambda_{1}y+4 \lambda_{1}^{3}t)}M_{22}(y,t,\lambda_{1}).
$$
Then,
$$
\overline{M_{22}(y,t,\lambda_{1})}=\frac{(\bar{\lambda}_{1}-\lambda_{1})^{2}}
{(\bar{\lambda}_{1}-\lambda_{1})^{2}+|C_{1}|^{2}e^{2i(\lambda_{1}y+4 \lambda_{1}^{3}t)}e^{-2i(\bar{\lambda}_{1}y+4 \bar{\lambda}_{1}^{3}t)}}.
$$
Substituting this result into the (\ref{4.3}), we get
\begin{align}\label{4.4}
M_{12}(y,t,\lambda)=\frac{\overline{C_{1}}(\bar{\lambda}_{1}-\lambda_{1})^{2}}
{(\lambda-\bar{\lambda}_{1})[(\bar{\lambda}_{1}-\lambda_{1})^{2}e^{2i(\bar{\lambda}_{1}y+4 \bar{\lambda}_{1}^{3}t)}+|C_{1}|^{2}e^{2i(\lambda_{1}y+4 \lambda_{1}^{3}t)}]}.
\end{align}
Let $\lambda_{1}=i\varepsilon$ , $\varepsilon>0$ and, In order to conveniently study the properties of the one soliton solution, we choose
$C_{1}=\pm 2i\varepsilon$. When $C_{1}=- 2i\varepsilon$,
substituting both  parameters into the (\ref{4.4}), it
comes into being
\begin{align}\label{4.5}
M_{12}(y,t,\lambda)=\frac{2i\varepsilon e^{-2(\varepsilon y-4\varepsilon^{3}t)} }
{(\lambda+i\varepsilon)[1-e^{-4(\varepsilon y-4\varepsilon^{3}t)}]}.
\end{align}
Then,
\begin{align*}
-i\lim_{\lambda \to \infty} (\lambda M(y,t, \lambda))_{12}=-(arctanh e^{-2(\varepsilon y-4\varepsilon^{3}t)})_{y}.
\end{align*}
where the $arctanh x$ is
the inverse function of $\tanh x$.
Furthermore,
\begin{align}\label{4.6}
&\int_{y}^{\infty}m(y',t)dy'=-i\int_{y}^{\infty}\lim_{\lambda \to \infty} (\lambda M(y',t, \lambda))_{12}dy'
\nonumber\\
=&-\int_{y}^{\infty} (arctanh e^{-2(\varepsilon y-4\varepsilon^{3}t)})_{y'}dy'
\nonumber\\
=&arctanh e^{-2(\varepsilon y-4\varepsilon^{3}t)}.
\end{align}
 The solution $q(x,t)$ in (\ref{3.7}) can transforms into
\begin{align}\label{4.7}
q(x,t)=e^{8arctanh e^{-2(\varepsilon y-4\varepsilon^{3}t)}}-1.
\end{align}
Let $\alpha (y,t)=e^{arctanh e^{-2(\varepsilon y-4\varepsilon^{3}t)}}$, we find that
$Ln \alpha (y,t)=arctanh e^{-2(\varepsilon y-4\varepsilon^{3}t)}$, then
$$
\tanh (ln \alpha (y,t))=e^{-2(\varepsilon y-4\varepsilon^{3}t)}
$$
i.e.
$$
\frac{e^{ln \alpha (y,t)}-e^{-ln \alpha (y,t)}}{e^{ln \alpha (y,t)}+e^{-ln \alpha (y,t)}}=e^{-2(\varepsilon y-4\varepsilon^{3}t)},
$$
we deduce
$$
\alpha^{2} (y,t)=-\tanh^{-1} (-\varepsilon y+4\varepsilon^{3}t).
$$
Equation (\ref{4.7}) can be written as
\begin{align}\label{4.8}
q(x,t)=(e^{arctanh e^{-2(\varepsilon y-4\varepsilon^{3}t)}})^{8}-1
=\tanh^{-4} (-\varepsilon y+4\varepsilon^{3}t)-1.
\end{align}
Substituting $y$ with $x$, (\ref{4.8}) becomes

\begin{align}\label{4.9}
q(x,t)=\tanh^{-4} (-\varepsilon x+4\varepsilon^{3}t-\varepsilon\gamma(x,t))-1
\end{align}
where $\gamma(x,t)=\int^{\infty}_{x}(1-\rho(\xi,t))d\xi$,
$\rho(x,t)=\tanh^{-2} (-\varepsilon x+4\varepsilon^{3}t-\varepsilon \gamma(x,t))$.
Then (\ref{4.9}) can be varied as
$(1+q(x,t))^{\frac{1}{2}}-1=\cosh^{2}(-\varepsilon x+4\varepsilon^{3}t-\varepsilon\gamma(x,t)),
$
hence the one soltion solution $q(x,t)$ has a singularity at the peak of the soliton so called
cusp soliton.

When $\lambda_{1}=i\varepsilon$ and $C_{1}=2i\varepsilon$,  the corresponding one soliton solution
$q(x,t)$ of (\ref{1.1}) can be expressed

\begin{align}\label{4.10}
q(x,t)=\tanh^{-4} (-\varepsilon x+4\varepsilon^{3}t-\varepsilon\gamma(x,t))-1
\end{align}
where $\gamma(x,t)=\int^{\infty}_{x}(1-\rho(\xi,t))d\xi$,
$\rho(x,t)=\tanh^{-2} (-\varepsilon x+4\varepsilon^{3}t-\varepsilon \gamma(x,t))$.

{\bf 4.2  Remark} In this paper, we use the uniform method to obtain the solution $q(x,t)$
of equation (\ref{1.1}) expressed by the (\ref{4.9}) and (\ref{4.10}). While  the \cite{W} applies the inverse
scatting method to get the solution $q(x,t)$.
If $\varepsilon=\kappa$($\kappa$ in the \cite{W}, To the one soliton solution, when $C_{1}=-2i\varepsilon$, expression of the solution in both paper is similar,
identically with $-\varepsilon x+4\varepsilon^{3}t$ in the
$$\tanh^{-4} (-\varepsilon x+4\varepsilon^{3}t-\varepsilon\gamma(x,t))$$
and  $\kappa x-4\kappa^{3}t$ in the $\tanh^{-4}(\kappa x-4\kappa^{3}t-\kappa x_{0}+\varepsilon_{+})$ in \cite{W}. There is different point that
the expression of one soliton solution in the two papers, one is dependent of the  $-\varepsilon\gamma(x,t)$ of $x$,
the other is $-\kappa x_{0}+\varepsilon_{+}$ of $x$.

\section*{Acknowledgments}

This work was supported
by grants from the National Science Foundation of China (Project
No.11271079), Doctoral Programs
Foundation of the Ministry of Education of China.

\vspace{0.8cm}

\renewcommand{\baselinestretch}{1.2}

\end{document}